\documentclass[12pt]{iopart}

\usepackage{iopams}
\usepackage[T1]{fontenc} 
\usepackage{graphicx}
\usepackage[usenames]{color}
\begin{document}

\title[]{Mobile vector  soliton in a spin-orbit coupled 
         spin-$1$ condensate}

\author{Sandeep Gautam and 
S. K. Adhikari\footnote{URL  http://www.ift.unesp.br/users/adhikari}}
\address{Instituto de F\'{\i}sica Te\'orica, Universidade Estadual
             Paulista - UNESP, \\ 01.140-070 S\~ao Paulo, S\~ao Paulo, Brazil}
\ead{sandeepgautam24@gmail.com and adhikari44@yahoo.com}
\vspace{10pt}
\begin{indented}
\item[] 4 February 2015
\end{indented}
      

\date{\today}
\begin{abstract}
We study the formation of bound states and three-component bright vector solitons in a 
quasi-one-dimensional spin-orbit-coupled hyperfine spin $f=1$ Bose-Einstein 
condensate using numerical solution and variational approximation of a 
mean-field model. In the antiferromagnetic domain, the solutions are 
time-reversal symmetric, and the component densities have multi-peak structure. 
In the ferromagnetic domain, the solutions violate time-reversal symmetry, 
and the component densities have single-peak structure. The dynamics of the 
system is not Galelian invariant. From an analysis of Galelian invariance, 
we establish that the single-peak ferromagnetic vector solitons are true 
solitons and  can move maintaining constant component densities, whereas the  
antiferromagnetic solitons cannot move with constant component densities. 
\end{abstract}

\maketitle

\section{Introduction}
\label{Sec-I}
Bright soliton is a self-reinforcing solitary wave that can traverse at a 
constant velocity without changing its shape due to a cancellation 
of the non-linear and dispersive interactions. The various systems
in which solitons have been studied include water waves, non-linear optics, 
Bose-Einstein condensates (BECs), etc. \cite{Kivshar}. Solitons have been observed  by manipulating the non-linear interaction near a 
Feshbach resonance \cite{Inouye} in a BEC of $^7$Li \cite{li} and $^{85}$Rb \cite{rb}. 
Solitons have also been studied in  binary 
BECs  \cite{Perez-Garcia}.  

 In a neutral spinor BEC with a nonzero hyperfine spin $f$, there is no spin-orbit (SO) coupling   between the   spin  of the atoms and
their center-of-mass motion \cite{stringari}. 
However, a synthetic SO coupling can be realized in a spinor BEC by controlling the atom-light interaction
leading to the generation of artificial Abelian and non-Abelian gauge potentials coupled
to the atoms \cite{Dalibard}.
 Solitons have been 
extensively studied in spinor BECs without SO coupling \cite{Ieda}. 
An SO coupling with equal Rashba \cite{Rashba} and 
Dresselhaus \cite{Dresselhaus} strengths was realized experimentally by Raman
dressing two atomic spin states with a pair of lasers \cite{Lin}. In that study, the SO 
coupling between 
two of the three spin components of the  
$f = 1$ state  5S$_{1/2}$ of $^{87}$Rb
$-$ the so-called  pseudospin-$1/2$ state $-$ was considered. 
There 
are other experimental studies on SO-coupled spinor BECs \cite{Aidelsburger}.
Solitonic structures in SO-coupled pseudospin-$1/2$ \cite{rela,Xu} and spin-$1$ BECs 
\cite{Liu} have also been investigated theoretically.

 In this letter, we study two types of three-component vector solitons in 
an SO-coupled spin-$1$ BEC in a
quasi-one-dimensional (quasi-1D) trap \cite{Salasnich} with  
{\em muliti-peak} or {\em single-peak} structure using a mean-field  coupled Gross-Pitaevskii (GP)
equation. 
A spin-$1$ spinor BEC is characterized by two interaction strengths,
namely $c_0\propto (a_0+2a_2)/3$ and $c_2 \propto (a_2-a_0)/3$, where $a_0$ and 
$a_2$ are $s$-wave scattering lengths in total spin $f_{\rm tot} = 0$ and 
2 channels respectively \cite{Ohmi}. For $c_2>0$ {(antiferromagnetic)}
the multi-peak structure emerges, whereas for $c_2<0$ {(ferromagnetic)}
the single-peak structure emerges.
We use variational method to determine the bright soliton solutions for 
the SO-coupled trapless BEC in each of the two domains. The appropriate 
variational {\em ansatz} in each of the domains is constructed using the 
solutions of the SO-coupled single particle Hamiltonian. {The variational analysis
provides  the necessary and sufficient conditions which $c_0$ and $c_2$
must satisfy to obtain a stable bright soliton.} We also compare the 
variational results with the  numerical solution of the 
GP equation.
 
 In Ref. 
\cite{Liu}, only antiferromagnetic { multi-peak} solitons for $c_2>0$ 
were 
identified as the bright solitons in a three-component  spin-$1$ SO-coupled BEC.
These solitons are time-reversal symmetric, but are not true   vector solitons as they cannot propagate maintaining the shape of the
individual  components.  
We demonstrate  that this  system can also support ferromagnetic { single-peak} solitons 
 for $c_2<0$, {  provided that $c_0+c_2<0$.
These solitons break the time-reversal symmetry of the Hamiltonian. Nevertheless,  they are 
shown to be true vector solitons as they can propagate with a constant velocity maintaining the shape
of the individual components.}


\section{Spin-Orbit-coupled BEC in quasi-1D trap}
\label{Sec-II}

We consider an SO-coupled spinor condensate in a quasi-1D trap in which the
trapping frequencies along the $y$ and $z$ axes ($\omega_y$ and $\omega_z$)  are much larger than that 
along the $x$ axis ($\omega_x$) \cite{Salasnich}. The single particle Hamiltonian of the 
condensate with equal strengths of Rashba \cite{Rashba} and Dresselhaus 
\cite{Dresselhaus} SO couplings in such a quasi-1D trap is \cite{H_zhai} 
\begin{equation}
H_0 = \frac{p_x^2}{2m} + V(x) + \gamma p_x \Sigma_x,
\label{sph} 
\end{equation}
where $p_x = -i\hbar\partial/\partial x$ is the momentum operator along x
axis, $V(x)=m\omega_x^2x^2/2$ is the harmonic trapping potential along $x$ 
axis, and $\Sigma_x$ is the irreducible representation of the $x$ component of 
the spin matrix:
\begin{eqnarray}
\Sigma_x= \frac{1}{\sqrt{2}}\left( \begin{array}
 {ccccc}
0 & 1 & 0\\
1 & 0 & 1\\
0 & 1 & 0 \end{array} \right).
\end{eqnarray}
This SO-coupling is distinct from a previous coupling \cite{gautam-1,gautam-2}
used in the study of a quasi-1D 
BEC.

Using the single particle model Hamiltonian  (\ref{sph}) and considering 
interactions in the  Hartree approximation, a quasi-1D \cite{Salasnich} spin-1 BEC 
can be described by the following set of three coupled mean-field partial 
differential equations for the wave-function components $\psi_j$ 
\cite{Ohmi,Kawaguchi}
\begin{eqnarray}
 i\hbar&\frac{\partial \psi_{\pm 1}}{\partial t} =
 \left( -\frac{\hbar^2}{2m}\frac{\partial^2}{\partial x^2}
 +V(x)+c_0\rho\right)\psi_{\pm 1}-\frac{i\hbar\gamma}{\sqrt{2}}
  \frac{\partial\psi_0}{\partial x}\nonumber\\ &
 +c_2(\rho_{\pm 1}+\rho_0-\rho_{\mp 1})\psi_{\pm 1}+ c_2 \psi_0^2\psi_{\mp 1}^*, \label{gp-1}\\
i\hbar &\frac{\partial \psi_0}{\partial t} = 
 \left( -\frac{\hbar^2}{2m}\frac{\partial^2}{\partial x^2}
 +V(x)+c_0\rho\right)\psi_0 -\frac{i\hbar\gamma}{\sqrt{2}}\Bigg(\frac{\partial\psi_{1}}{\partial x}  \nonumber\\
  &+\frac{\partial\psi_{-1}}{\partial x}\Bigg) 
  +c_2(\rho_1+\rho_{-1})\psi_0+ 2c_2\psi_0^*\psi_1\psi_{-1},\label{gp-3}
\end{eqnarray}
where $c_0 = 2\hbar^2 (a_0+2a_2)/(3m l_{yz}^2)$, 
$c_2 = 2\hbar^2 (a_2-a_0)/(3m l_{yz}^2)$,   
$a_0$ and $a_2$ are the $s$-wave scattering lengths in the total 
spin $f_{\mathrm{tot}} = 0$ and $2$ channels, respectively,    
$\rho_j = |\psi_j|^2$ with $j=1,0,-1$ are the component densities, 
$\rho(x)= \sum_{j=-1}^1\rho_j$ is the total density, 
and $l_{yz} = \sqrt{\hbar/(m\omega_{yz})}$ with  $ \omega_{yz}=\sqrt{\omega_y\omega_z}$
is the oscillator length in the transverse $y-z$ plane.
For the sake of simplicity, let us transform   (\ref{gp-1})-(\ref{gp-3})
into dimensionless form using 
\begin{equation}
 \tilde{t} = \omega_x t,~\tilde{x} = \frac{x}{l_0},
 ~\phi_j(\tilde{x},\tilde{t}) = 
 \frac{\sqrt{l_0}}{\sqrt{N}}\psi_j(\tilde{x},\tilde{t}), 
\end{equation}
where $l_0=\sqrt{\hbar/(m\omega_{x}}$) is the oscillator length along $x$ axis,
and $N$ is the total number of atoms:
\begin{eqnarray}
 i &\frac{\partial \phi_{\pm 1}}{\partial \tilde{t}} =
 \left( -\frac{1}{2}\frac{\partial^2}{\partial \tilde{x}^2}
 +\tilde{V}(\tilde{x})+\tilde{c}_0\tilde{\rho}\right)\phi_{\pm 1}-\frac{i\tilde{\gamma}}{\sqrt{2}}
  \frac{\partial\phi_0}{\partial \tilde{x}}\nonumber\\ 
 &+c_2(\tilde{\rho}_{\pm 1}+\tilde{\rho}_0-\tilde{\rho}_{\mp 1})\phi_{\pm 1}+ c_2 \phi_0^2\phi_{\mp 1}^*,\label{gps-1}\\
 i&\frac{\partial \phi_0}{\partial \tilde{t}} = 
 \left( -\frac{1}{2}\frac{\partial^2}{\partial \tilde{x}^2}
 +\tilde{V}(\tilde{x})+\tilde{c}_0\tilde{\rho}\right)\phi_0 -
\frac{i\tilde{\gamma}}{\sqrt{2}}\Bigg(\frac{\partial\phi_{1}}{\partial \tilde{x}}  \nonumber\\
  &+\frac{\partial\phi_{-1}}{\partial \tilde{x}}\Bigg) 
  +c_2(\tilde{\rho}_1+\tilde{\rho}_{-1})\phi_0+ 2\tilde{c}_2\phi_0^*\phi_1\phi_{-1},\label{gps-3},
\end{eqnarray}
where $\tilde{V} = \tilde{x}^2/2$, 
$\tilde{\gamma} = \hbar k_r/(m\omega_x l_0)$, 
$\tilde{c}_0 = 2N (a_0+2a_2)l_0/(3l^2 _{yz})$,
$\tilde{c}_2 = 2N (a_2-a_2)l_0/(3l^2_{yz})$,
$\tilde{\rho}_j = |\phi_j|^2$ with $j=1,0,-1$, 
and $\tilde{\rho} = \sum_{j=-1}^1|\phi_j|^2.$
The total density is now normalized to unity, i.e., 
$
 \int_{-\infty}^{\infty} \tilde{\rho}(\tilde{x})d\tilde{x} = 1.
$
We present the scaled variables without tildes in the rest of
the letter for notational simplicity. For a non-interacting trapless
system [$V(x)=c_0=c_2=0$], there are two linearly independent solutions
of the SO-coupled set of equations  (\ref{gps-1})-(\ref{gps-3}) with the lowest energy
 $E_{\rm min} = -N\gamma^2/2$:
\begin{equation}
\Phi_1=  \frac{e^{i\gamma x}}{2}\left( \begin{array}{c}
1 \\
-\sqrt{2}\\
1 \end{array} \right), \quad\\ 
\Phi_2= \frac{e^{-i\gamma x}}{2}\left( \begin{array}
 {c}
1 \\
\sqrt{2}\\
1 \end{array} \right),
\label{spinors}
\end{equation}
where wave functions $\Phi_1$ and $\Phi_2$ are normalized to unity.   Hence, the most general
solution of Eqs. (\ref{gps-1})-(\ref{gps-3}) for a non-interacting trapless system
with a fixed density $n$ is given by the linear superposition of $\sqrt{n}\Phi_1$
and $\sqrt{n}\Phi_2$
\begin{eqnarray}
\sqrt{n}\Phi & \equiv &
\left( \begin{array}{c}
\phi_{1}\\
\phi_{0}\\
\phi_{-1}\end{array} \right) = \sqrt{n}(\alpha_1 \Phi_1 + \alpha_2 \Phi_2) ,\nonumber\\
    &=& \frac{\sqrt{n}}{2}\left( \begin{array}{c}
\alpha_1e^{i\gamma x} + \alpha_2e^{-i\gamma x}\\
-\sqrt{2}\alpha_1e^{i\gamma x} +\sqrt{2} \alpha_2e^{-i\gamma x}\\
\alpha_1e^{i\gamma x} + \alpha_2e^{-i\gamma x}\end{array}\right),
\label{single_part_sol}
\end{eqnarray}
where $|\alpha_1|^2+|\alpha_2|^2=1$ to ensure that $\Phi$ is  normalized to unity.

The energy of the BEC in scaled units is
\begin{eqnarray}
 E &=  N\int_{-\infty}^{\infty} \Bigg\{\frac{1}{2}\sum_{j=-1}^1 \left|\frac{d\phi_j}{dx}\right|^2
  -\frac{i\gamma}{\sqrt{2}} \left(\phi_1^*+\phi_{-1}^*\right)\frac{d\phi_0}{dx}\nonumber\\
  &-\frac{i\gamma}{\sqrt{2}}\phi_0^*\left(\frac{d\phi_1}{dx} + \frac{d\phi_{-1}}{dx}\right)
  +  \frac{{c_0}\rho^2 + {c_2}|\mathbf F|^2}{2}
 \Bigg\}dx,
  \label{energy}
\end{eqnarray}
where $\mathbf F$ is spin density vector, whose three components $F_x,F_y$, and $F_z$
are defined as
\begin{eqnarray}
F_x &=& \frac{1}{\sqrt{2}}\left[\phi_0\left(\phi_1^*+\phi_{-1}^*\right)+\phi_0^*(\phi_1+\phi_{-1})\right],\\
F_y &=& \frac{i}{\sqrt{2}}\left[\phi_0\left(\phi_{-1}^*-\phi_1^*\right)+\phi_0^*(\phi_1-\phi_{-1})\right],\\
F_z &=& |\phi_1|^2-|\phi_{-1}|^2.
\end{eqnarray}
Hence, for the SO-coupled Hamiltonian with its general solution given by  (\ref{single_part_sol}), we get
\begin{eqnarray}
        F_x&=& n(|\alpha_2|^2 -|\alpha_1|^2),\\
        F_y   &=& F_z=0.
\end{eqnarray}
Also, the magnetization ${\cal M} = \int F_z dz = 0$ for
minimum energy solutions of the single-particle SO-coupled Hamiltonian.

Now, let us switch on the interactions; the interaction energy per 
particle for the uniform system is \cite{Kawaguchi}
\begin{eqnarray}
\epsilon_{\rm int} &=&\left[\frac{c_0}{2}n + \frac{c_2}{2n}|\mathbf F|^2\right],\nonumber\\  
                   &=&\left[\frac{c_0}{2}n +\frac{c_2}{2}n\left(|\alpha_2|^2 -|\alpha_1|^2\right)^2\right]. 
\end{eqnarray}
If $c_2>0$, then the BEC is in the antiferromagnetic or polar phase, and the minimum of $\epsilon_{\rm int}$ 
corresponds to $|\alpha_1| = |\alpha_2| = 1/\sqrt{2}$ leading to $|\mathbf F|/n = 0$. 
In this case, the wave function (\ref{single_part_sol}) is time-reversal
symmetric.
On the other hand, for $c_2<0$, the BEC is in the ferromagnetic phase, and  $\epsilon_{\rm int}$ can be minimized if $|\alpha_1| = 1, 
|\alpha_2| = 0$ or $|\alpha_1| = 0, |\alpha_2| = 1$, which leads to $|\mathbf F|/n =1$. 
This corresponds to the wave functions (\ref{spinors}) apart from a multiplying phase factor. 
These states are degenerate, violate the time-reversal symmetry and are mutually connected by the time-reversal operator.
These are the only two distinct structures which emerge as the ground
states in the SO-coupled quasi-1D BECs.  
In a quasi-two-dimensional BEC with Rashba or Dresselhaus SO coupling, there is a circular degeneracy in
the energy eigen functions of the single particle Hamiltonian \cite{zhai}. Hence, depending
upon the interaction parameters, more than two plane waves can also 
superpose resulting in different types of lattice 
structures in ground state density profiles \cite{Kawakami}.

\section{Bright solitons}
\label{Sec-III}

\subsection{Stationary bright solitons}
\label{Sec-III-A}

Stationary bright solitons can emerge as the ground state of a spinor
BEC with attractive interactions \cite{rela,Liu}. We use variational method to determine
the bright soliton solutions of  (\ref{gps-1})-(\ref{gps-3}).  As has been discussed in 
Sec. \ref{Sec-II}, an SO-coupled spinor BEC can have two
types of ground states depending upon the sign of  $c_2$.
This necessitates the use of two different variational {\em ansatz}
in these two domains.

{\em Antiferromagnetic phase} ($c_2>0$): Here we consider the following variational {\em ansatz}
to determine the shape of the soliton 
\begin{eqnarray}
\Phi_{\rm var} &=& \frac{\sqrt{\sigma}}{2}\left( \begin{array}{c}
\pm \cos(\gamma x) \\
\mp \sqrt{2}i\sin(\gamma x)\\
\pm \cos(\gamma x)\\ \end{array} \right){\rm sech}(\sigma x),
\label{ansatz_1}
\end{eqnarray}
where $\sigma$ is a variational parameter and characterizes the width and the strength of
the bright soliton.
The {\em ansatz}  (\ref{ansatz_1}) corresponds to the wave function (\ref{single_part_sol}) 
with $\alpha_1=\alpha_2= \pm 1/\sqrt{2}$ multiplied by the localized spatial soliton 
$\sqrt{\sigma/2} \mathrm{sech} (\sigma x)$ instead of $\sqrt{n}$.
As two solutions (\ref{spinors}) are degenerate, and a mixing between them is allowed,
the soliton profile could have a multi-peak structure. 
Noting that in the $c_2>0$ domain, for $|\alpha_1|=|\alpha_2| = 1/\sqrt{2}$, one can have other
choices for the variational {\em ansatz} like
\begin{eqnarray}
\Phi_{\rm var} &=& \frac{\sqrt{\sigma}}{2}\left( \begin{array}{c}
\pm i\sin(\gamma x) \\
\mp \sqrt{2}\cos(\gamma x)\\
\pm i\sin(\gamma x)\\ \end{array} \right){\rm sech}(\sigma x),
\label{ansatz_1a}\\
\Phi_{\rm var} &=& \frac{\sqrt{\sigma}}{4}\left( \begin{array}{c}
e^{i\gamma x}\pm i e^{-i \gamma  x} \\
-\sqrt{2}(e^{i\gamma x}\mp i e^{-i \gamma  x})\\
e^{i\gamma x}\pm i e^{-i \gamma  x}\\ \end{array} \right){\rm sech}(\sigma x),
\label{ansatz_1b}
\end{eqnarray}
etc, where  (\ref{ansatz_1a}) and (\ref{ansatz_1b}) correspond to 
$\alpha_1=-\alpha_2=\pm 1/\sqrt{2}$ and $\alpha_1=\mp i\alpha_2=1/\sqrt{2}$,
respectively, in  (\ref{single_part_sol}).
Substituting any of these {\em ansatz} in  (\ref{energy}), the energy
of the soliton is
\begin{equation}
E = \frac{N}{6} \left(-3 \gamma ^2+\sigma ^2+\sigma  c_0\right).
\end{equation} 
The minima of this energy occurs at
\begin{equation}
\sigma  = -\frac{c_0}{2},
\label{sw1}
\end{equation}
provided $c_0<0$. Hence, the SO-coupled spin-$1$ spinor BEC can support
an antiferromagnetic bright soliton defined by  (\ref{ansatz_1}) [or 
 (\ref{ansatz_1a}) or  (\ref{ansatz_1b})] and  (\ref{sw1}), provided that 
$c_0<0$ and $c_2>0$. From  (\ref{ansatz_1}) and (\ref{sw1}) it is evident
that the wavefunction of the bright soliton is independent of the strength
of spin-exchange interactions $c_2$. This is expected since for $c_2>0$, there is no
contribution to the energy from the $c_2$-dependent term of the SO-coupled spinor BEC. 

{\em Ferromagnetic phase} ($c_2<0$): Here we consider the following variational 
{\em ansatz} 
\begin{equation}
\Phi_{\rm var} = \frac{\sqrt{\sigma}e^{i\gamma x}}{2\sqrt{2}}\left( \begin{array}{c}
1 \\
-\sqrt{2}\\
1 \end{array} \right){\rm sech}(\sigma x),
\label{ansatz_2}
\end{equation}
where  $\sigma$ is, again, a variational parameter 
characterizing the width and the strength of the bright soliton. This variational {\em ansatz} 
corresponds to $\alpha_1 = 1, \alpha_2 = 0$ in  (\ref{single_part_sol}) multiplied 
by the localized bright soliton $\sqrt {\sigma/2} \mathrm{sech}(\sigma x)$ instead of $\sqrt{n}$. 
In this case the soliton will have a single peak.
Also, the {\em ansatz} like $-\Phi_{\rm var}, or \pm i\Phi_{\rm var}$ 
are equally reasonable choices  and correspond
to $\alpha_1 = -1, \alpha_2 = 0$ and $\alpha_1 = \pm i, \alpha_2 = 0$, respectively, in (\ref{single_part_sol}). 
Substituting  (\ref{ansatz_2})
in  (\ref{energy}), the energy of the soliton is
\begin{equation}
E = \frac{1}{6} \left(-3 \gamma ^2+\sigma ^2+\sigma  c_0+\sigma  c_2\right).
\end{equation}
The minima of this energy occurs at
\begin{equation}
\sigma = -\frac{1}{2} \left(c_0+c_2\right),
\label{sw2}
\end{equation}
provided $c_0+ c_2 < 0$. Hence the SO-coupled spinor  BEC can have a ferromagnetic
soliton defined by  (\ref{ansatz_2}) and (\ref{sw2}), provided
$c_2<0$ and $c_0+ c_2 < 0$. In this case, unlike in the case of an antiferromagnetic soliton, the bright soliton profile
is sensitive to both $c_0$ and $c_2$.

\subsection{Moving bright solitons}
\label{Sec-III-B}
If $\Phi$ is static bright 
solitonic solution of the coupled  equations (\ref{gps-1})-(\ref{gps-3}), then the 
Galilean invariance of these equations ensures that a soliton moving with 
velocity $v$ is defined as
\begin{equation}
\Phi_M (x,t) = \Phi(x-vt,t)e^{ivx+i(\sigma-v^2)t/2},
\end{equation}
where $\sigma$ characterizes the  width and the strength of the 
soliton.  The breakdown of the Galilean invariance of the SO-coupled equation  can be 
explicitly seen by using the transformation $x'=x+vt$, $t=t'$, 
where $v$ is the velocity of the unprimed coordinate system with respect to 
primed coordinate system, then the wavefunction $\Phi$ of 
 (\ref{gps-1})-(\ref{gps-3}) should transform to $\Phi_M$ as
\begin{equation}
\Phi(x,t) = \Phi_M(x',t')e^{-ivx'-i(\sigma-v^2)t'/2}.
\label{gs}
\end{equation}
Now, substituting  (\ref{gs}) in  (\ref{gps-1})-(\ref{gps-3}) and 
using $\partial/\partial x=\partial/\partial x'$
and $\partial/\partial t = \partial/\partial t'+v\partial/\partial x'$, 
we obtain
\begin{equation}
i\frac{\partial \Phi_M(x',t')}{\partial t'} = \left[-\frac{1}{2}\frac{\partial^2}{\partial x'^2}-
\gamma\Sigma_x\left(i\frac{\partial}{\partial x'}+v\right)\right]\Phi_M(x',t'),
\label{mov_sol}
\end{equation}
where the terms proportional to $c_0$ and $c_2$ have been suppressed for the
sake of simplicity in addition to a $\sigma$-dependent additive term which does 
not contribute to the dynamics. 
The presence of the extra term $-\gamma\Sigma_x v\Phi_M(x',t')$ on the right hand side of 
(\ref{mov_sol})
shows that the SO-coupled Hamiltonian is no longer Galilean invariant and the SO-coupled 
soliton solution of the GP equation will depend on its velocity $v$.
The  SO-coupled equation (\ref{mov_sol}), in the absence of trap and interactions, has the 
solutions $\Phi_1$ and $\Phi_2$ of  Eq. (\ref{spinors}) with energies  $E=-N(\gamma^2/2-\gamma v)$ 
and $E=-N(\gamma^2/2+\gamma v)$, respectively. For $v=0$, the two solutions (\ref{spinors})
were degenerate, and this degeneracy has been removed in the case of the SO-coupled moving solutions. 
In the antiferromagnetic phase, a multi-peak  solution was possible through a mixture of 
two degenerate solutions (\ref{spinors}) for $v=0$. For a nonzero $v$, the degeneracy is removed and 
such a mixing is not possible. This means that the multi-peak soliton cannot propagate with 
a constant velocity maintaining its shape and energy. For the moving multi-peak soliton profile, 
the variational analysis of Sec. \ref{Sec-III-A} will no longer be valid.
In the ferromagnetic phase, as a mixing between the two degenerate solutions 
is not allowed, one can only have a single-peak soliton which can propagate  
with a constant velocity maintaining its shape, and the variational analysis 
presented in Sec. \ref{Sec-III-A} remains valid.

\section{Results and conclusions}
\label{Sec-IV}
We numerically solve the coupled equations  (\ref{gps-1})-(\ref{gps-3}) using the
split-time-step Crank-Nicolson method \cite{Wang,Muruganandam} with real- and imaginary-time propagations. The ground 
state is determined by solving  (\ref{gps-1})-(\ref{gps-3}) using 
imaginary-time propagation, which neither conserves norm nor magnetization. 
Both norm and magnetization can be fixed by transforming the 
wave-function components as
\begin{equation}
 \phi_j(x,\tau+d\tau) = d_j \phi_j
(x,\tau),
\end{equation}
after each iteration in imaginary time $\tau = -it$, where $d_j$'s with 
$j=1,0,-1$ are the normalization constants. The $d_j$'s are defined as 
\cite{Bao, gautam-2}
\begin{eqnarray}
 d_0 &=\frac{\sqrt{1-{\cal M}^2}}{\sqrt{N_0+\sqrt{4(1-{\cal M}^2)
       N_1N_{-1}+{\cal M}^2N_0^2}}},\\
 d_1 &=\sqrt{\frac{1+{\cal M}-c_0^2N_0}{2N_1}},\\
 d_{-1} &=\sqrt{\frac{1-{\cal M}-c_0^2N_0}{2N_{-1}}},
\end{eqnarray}
and here $N_j = \int |\phi_j(x,\tau)|^2 dx$. These normalizations
ensure simultaneous conservation of norm and magnetization after each iteration
in imaginary time.
The
spatial and time steps used in the present work are
$\delta x = 0.05$ and $\delta t = 0.000125$, respectively.

\begin{figure}[!h]
\begin{center}
\includegraphics[trim = 0mm 0mm 2cm 0mm, clip,width=.4\linewidth,clip]{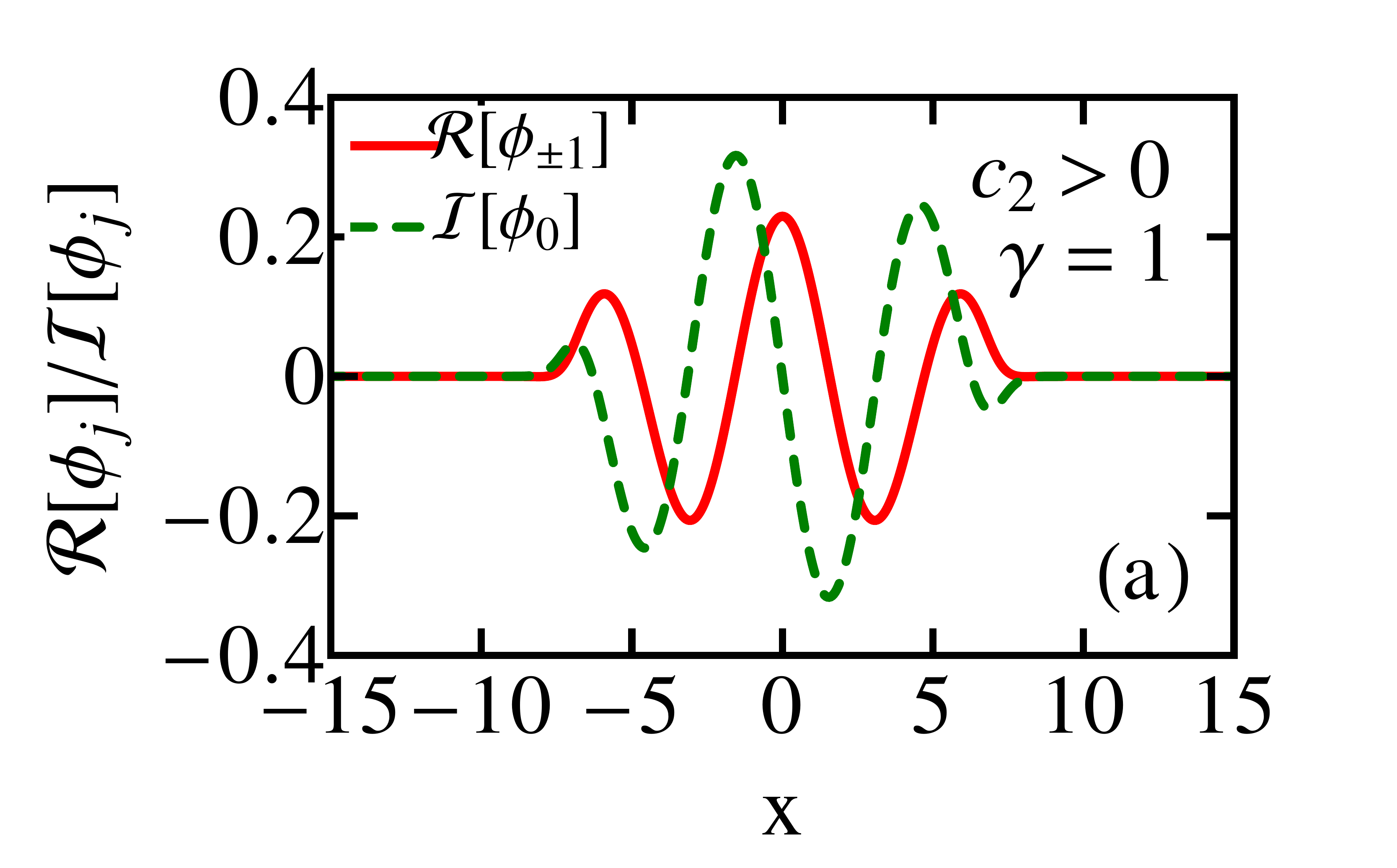}
\includegraphics[trim = 0mm 0mm 2cm 0mm, clip,width=.4\linewidth,clip]{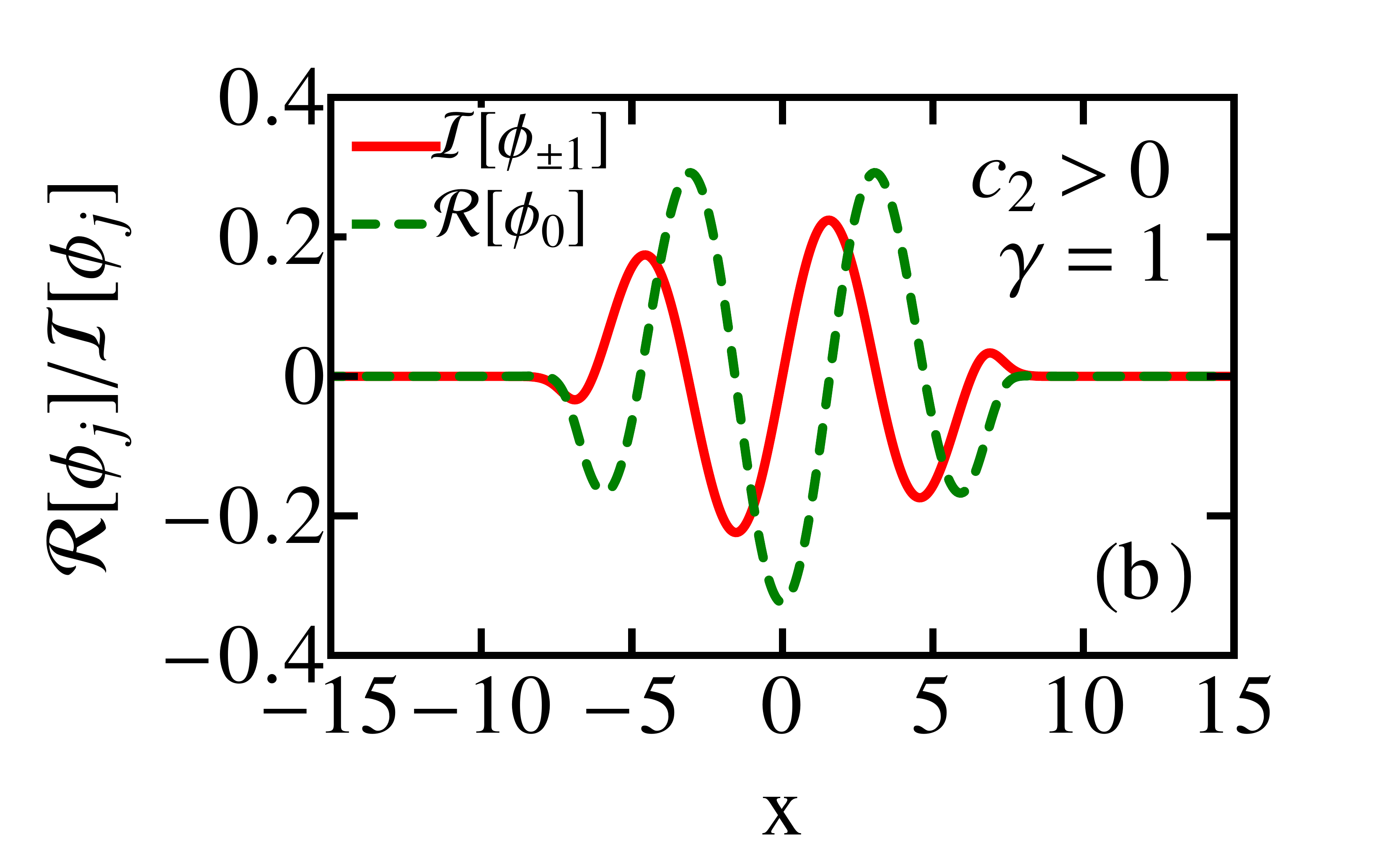}
\includegraphics[trim = 0mm 0mm 2cm 0mm, clip,width=.4\linewidth,clip]{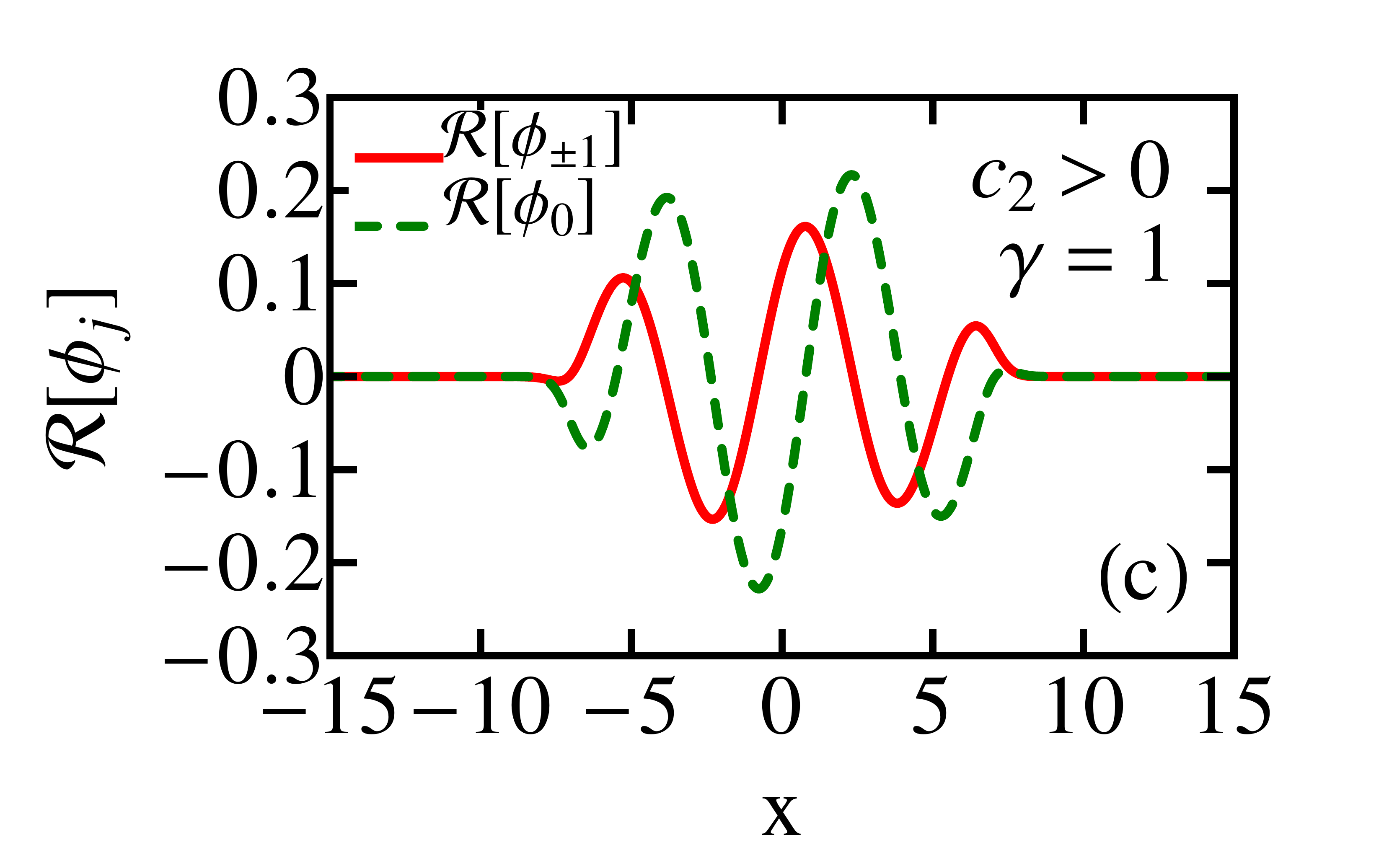}
\includegraphics[trim = 0mm 0mm 2cm 0mm, clip,width=.4\linewidth,clip]{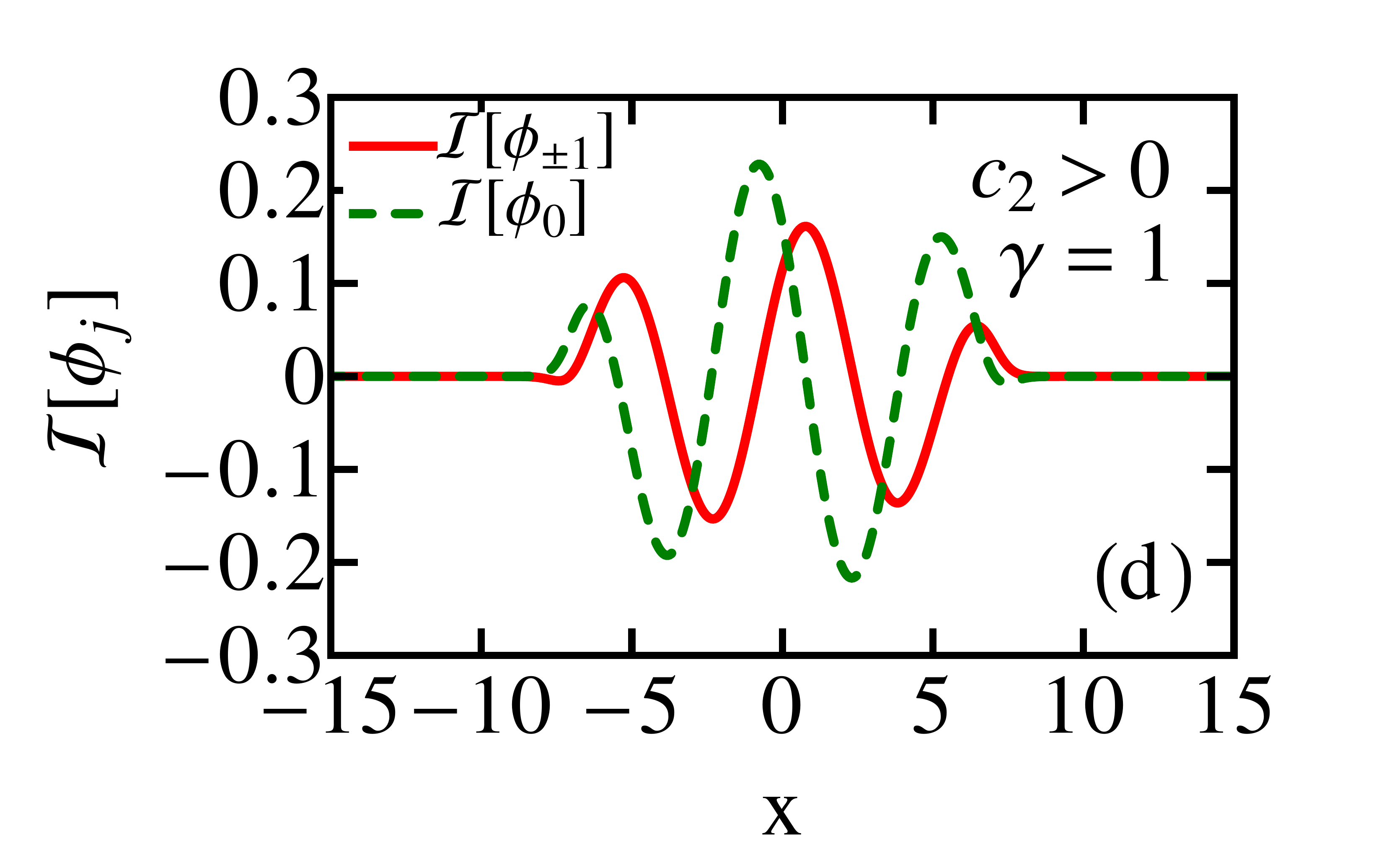}
\includegraphics[trim = 0mm 0mm 2cm 0mm, clip,width=.4\linewidth,clip]{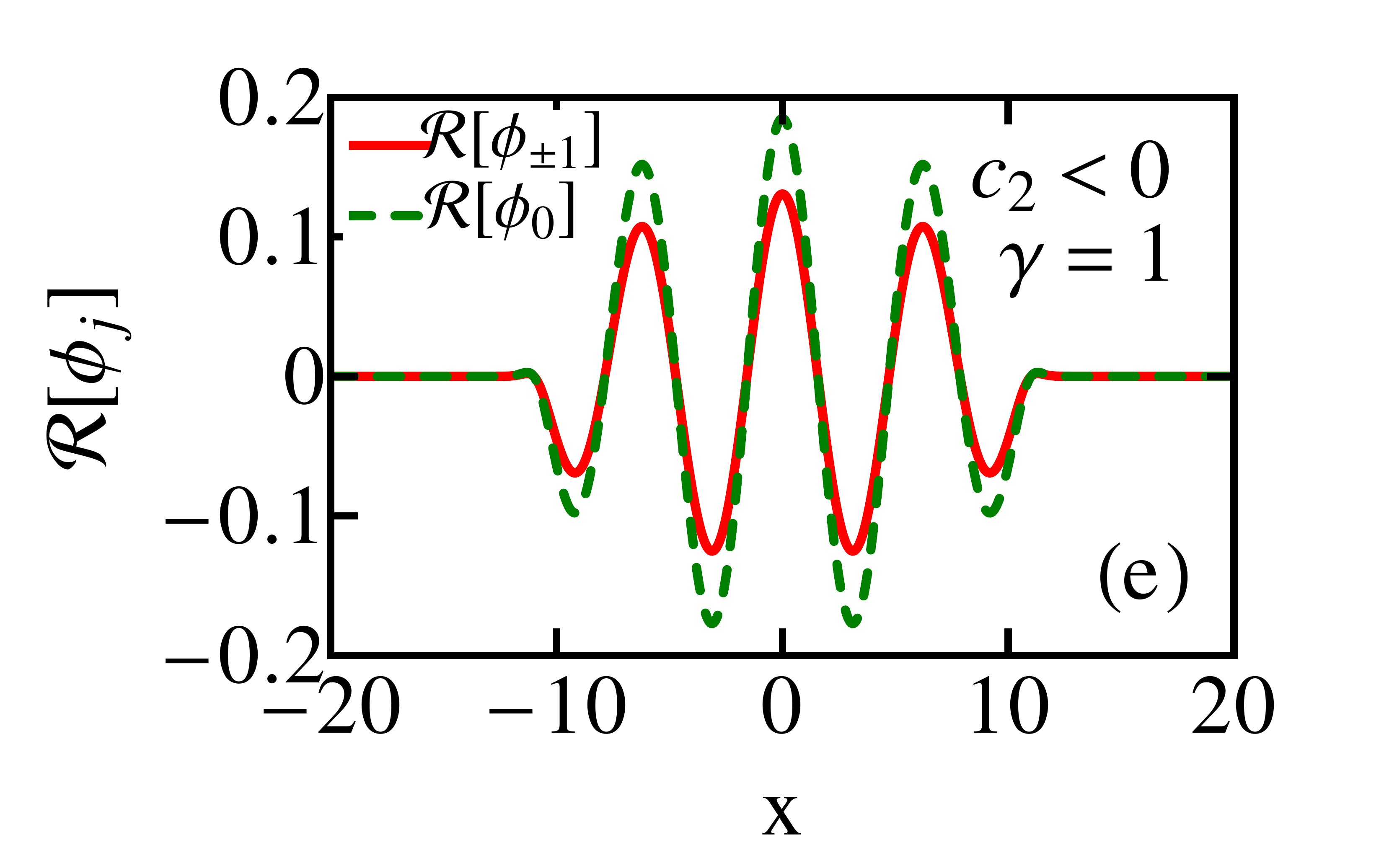}
\includegraphics[trim = 0mm 0mm 2cm 0mm, clip,width=.4\linewidth,clip]{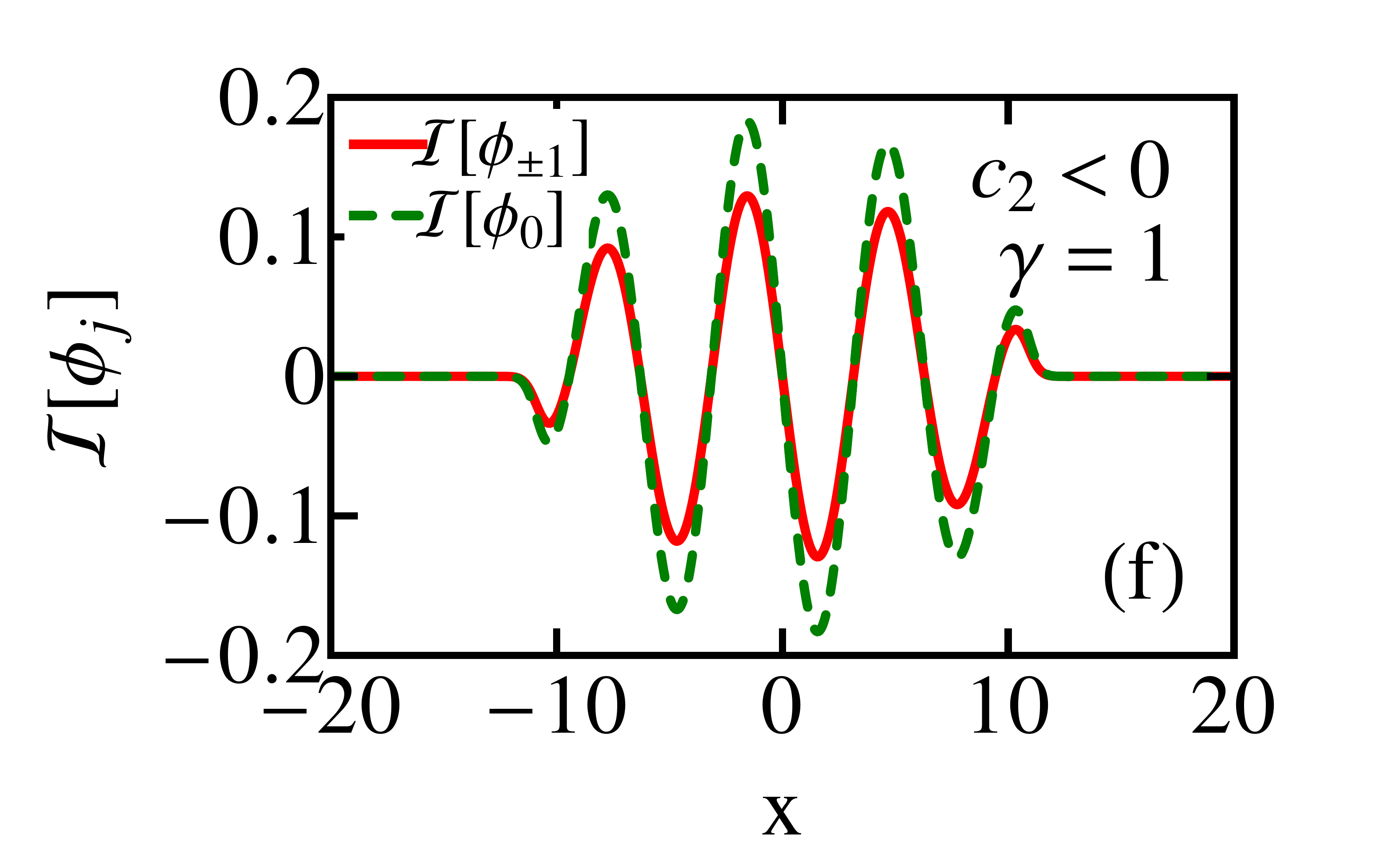}
\includegraphics[trim = 0mm 0mm 2cm 0mm, clip,width=.4\linewidth,clip]{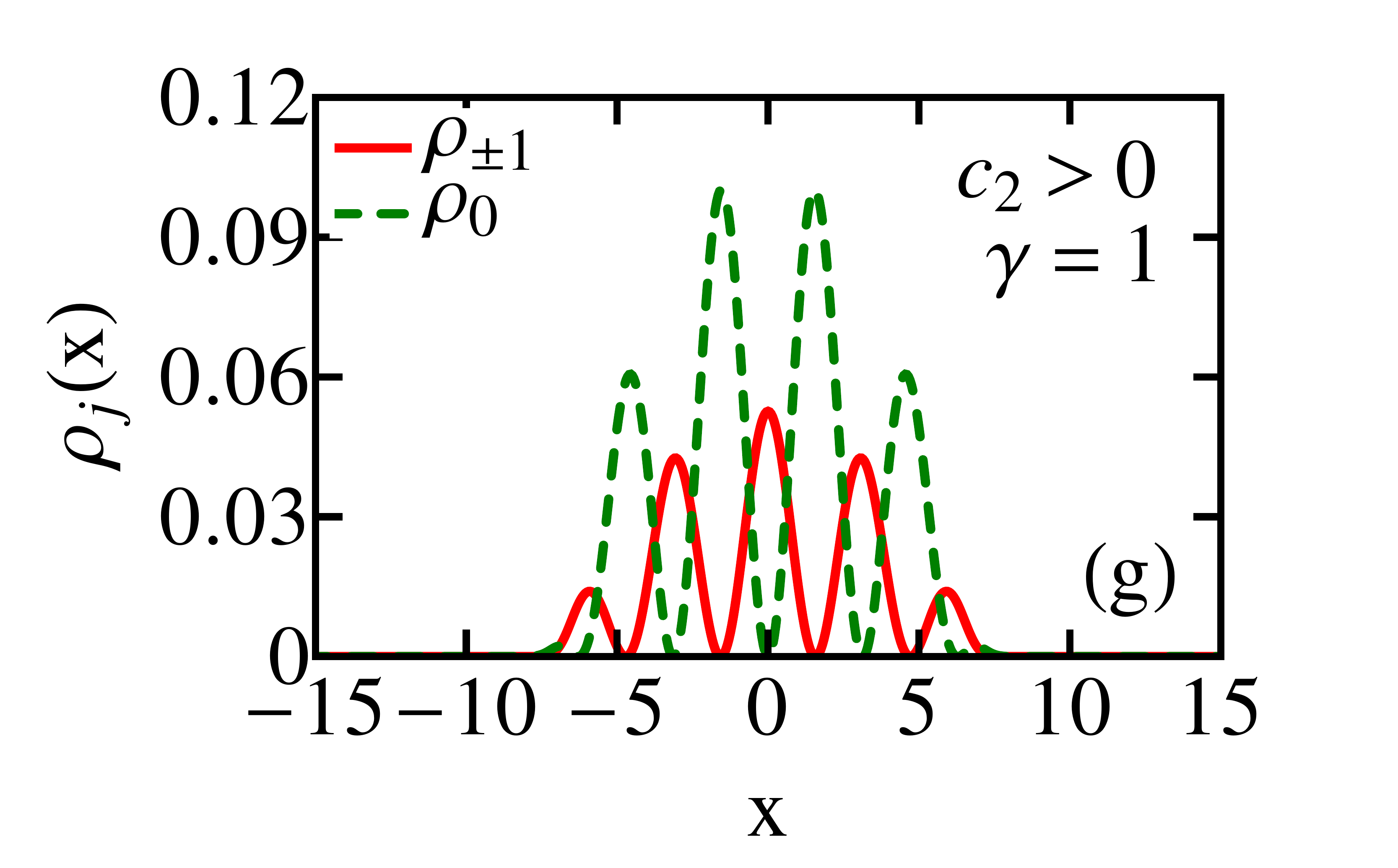}
\includegraphics[trim = 0mm 0mm 2cm 0mm, clip,width=.4\linewidth,clip]{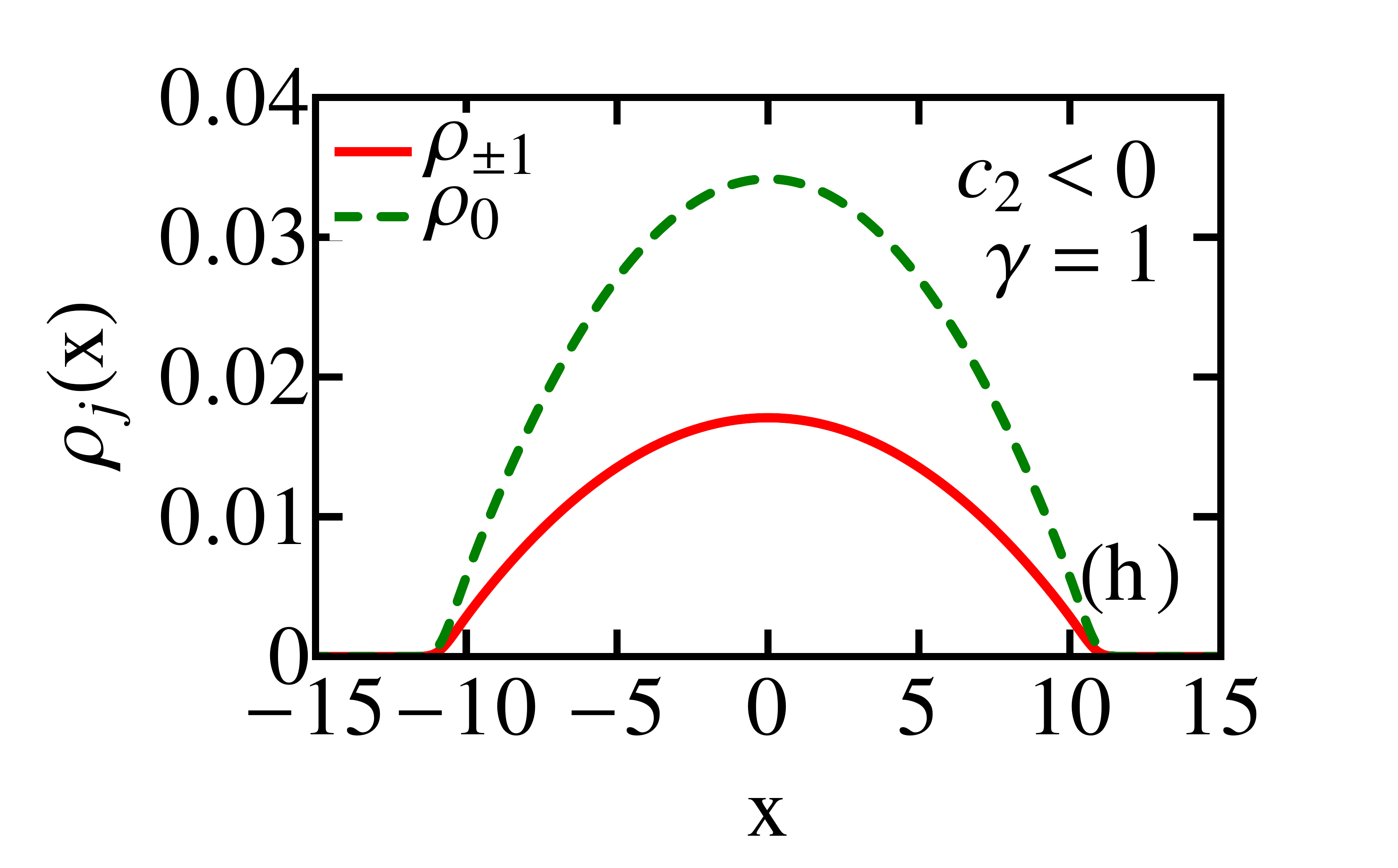}
\caption{(Color online) (a) and (b) show the  {\it non-zero} real (${\cal R}$) 
and imaginary (${\cal I}$) parts of wave-function components for the antiferromagnetic 
BEC of $^{23}$Na with $N = 10 000$, $c_0 = 241.28$, $c_2 = 7.76$, and $\gamma = 1$. 
For the same interaction parameters and SO coupling, (c) and (d) show the real and 
imaginary parts, respectively, of complex wave-function components for the antiferromagnetic BEC 
of $^{23}$Na. The three solutions correspond to 
different choices of $\alpha_1$ and $\alpha_2$ in  (\ref{initial_guess}). 
(e) and (f) show the real and imaginary parts, respectively, of complex wave-function components
for the ferromagnetic BEC of $^{87}$Rb 
with $N = 10 000$, $c_0 = 885.71$, $c_2 = -4.09$, and $\gamma = 1$. 
(g) shows the density distribution corresponding to  (a),(b), and (c) and (d),
whereas (h) shows the same corresponding to (e) and (f).
In this and the following   figures, all quantities are dimensionless.}
\label{fig-1} \end{center}
\end{figure}

\begin{figure}[!h]
\begin{center}
\includegraphics[trim = 0mm 0mm 3cm 0mm, clip,width=.4\linewidth,clip]{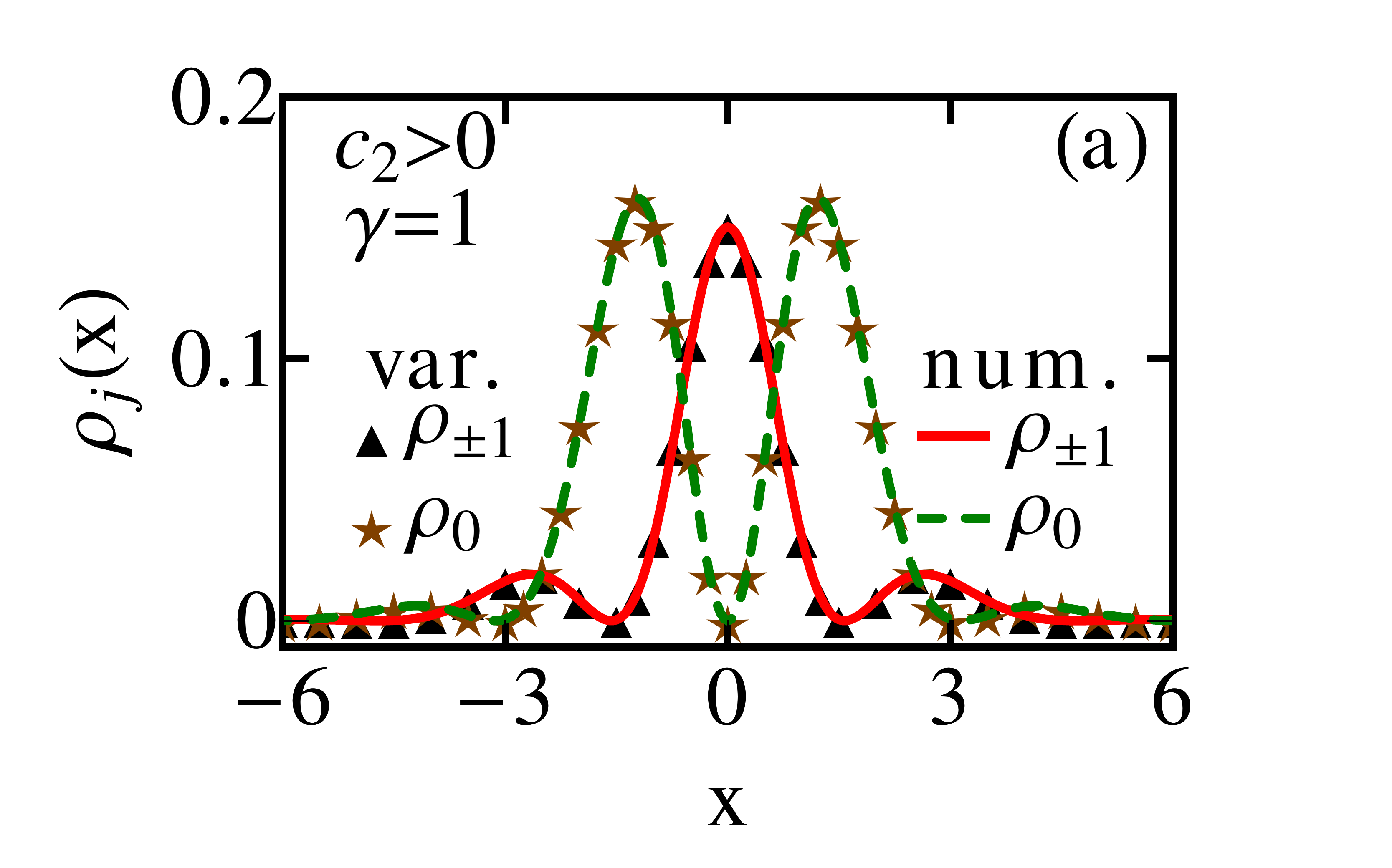}
\includegraphics[trim = 0mm 0mm 3cm 0mm, clip,width=.4\linewidth,clip]{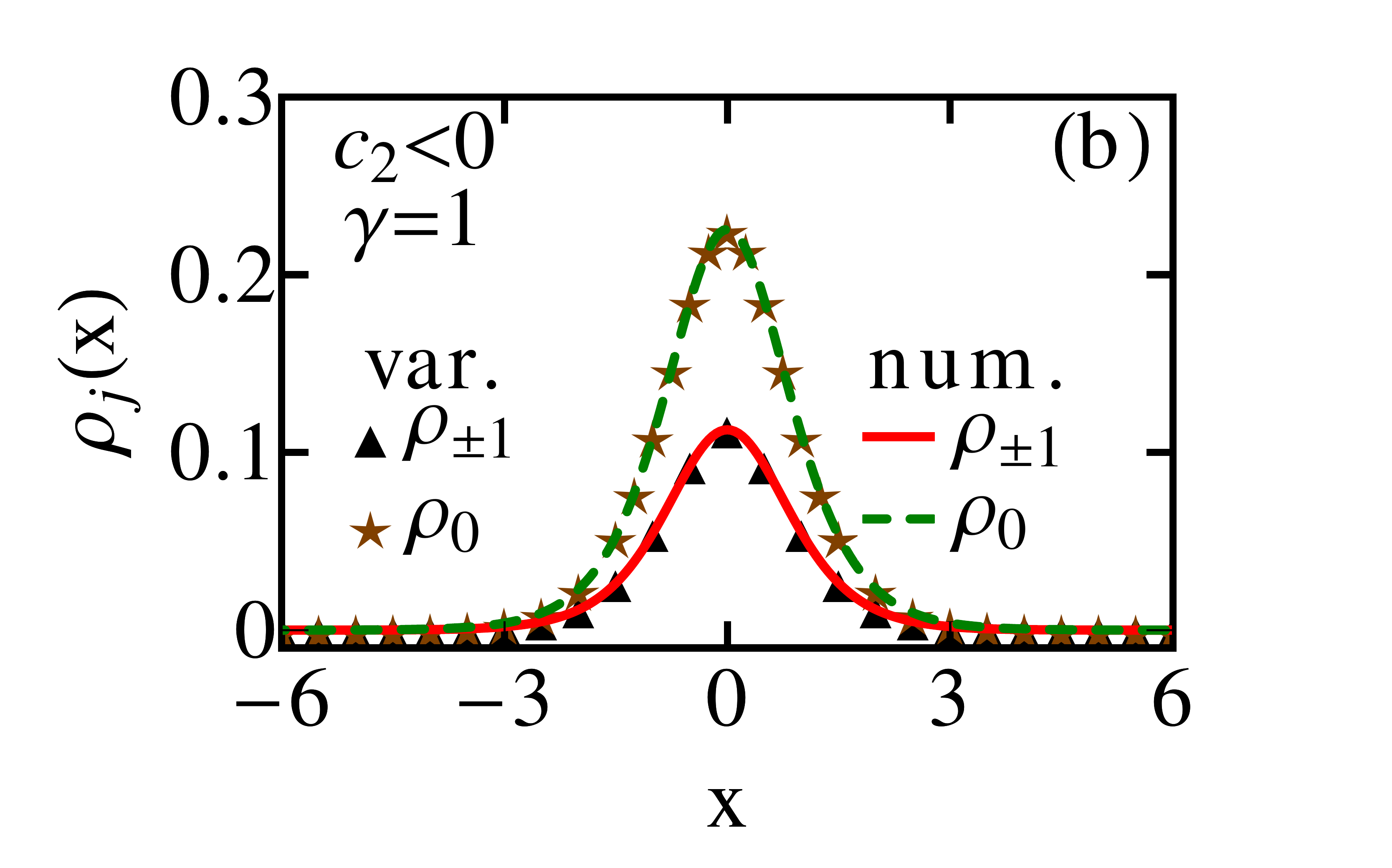}
\caption{(Color online) (a)  Numerical and variational  densities
of the  wave-function components for a bright soliton in antiferromagnetic
phase with $c_0 = -1.2$, $c_2 = 0.3$, and $\gamma = 1$.  (b) The same 
of ferromagnetic phase with $c_0 = -1.5$, $c_2 = -0.3$, and $\gamma = 1$.}
\label{fig-2} \end{center}
\end{figure}

We consider an SO-coupled spin-$1$ spinor BEC of $10000$ 
 $^{23}$Na or $^{87}$Rb atoms trapped in a harmonic trapping potential 
with $\omega_x/(2\pi) = 20$ Hz and $\omega_y/(2\pi) = \omega_z/(2\pi) =400$ Hz. The oscillator
lengths for $^{23}$Na with these parameters are $l_0 = 4.69~\mu$m
and $l_{yz} = 1.05~\mu$m, whereas those for $^{87}$Rb are $l_0 = 2.41~\mu$m
and $l_{yz} = 0.54~\mu$m. We use these values of $l_0$ for writing the dimensionless  
GP equations (\ref{gps-1})-(\ref{gps-3}) for the trapped states, whereas for
solitons $l_0 = 4.69~\mu$m in this letter.
 The scattering lengths of $^{23}$Na
in total spin $f_{\rm tot} = 0$ and $2$ channels are $a_0 = 2.646$ nm, $a_2 = 2.919$ nm, 
respectively \cite{Bao}, resulting in $c_0 = 241.28$ and $c_2 = 7.76$. Similarly, the scattering
lengths of $^{87}$Rb are $a_0 = 5.387$ nm and $a_2 = 5.313$ nm \cite{Bao}, 
leading to $c_0 = 885.71$ and $c_2 = -4.09$. In imaginary time propagation,
we use a real Gaussian function multiplied by the solution of the 
single-particle 
SO-coupled Hamiltonian as the initial input for the component wavefunctions, i.e.,
\begin{equation}
\Phi_{initial} = \frac{e^{-x^2/2}}{2\sqrt{\sqrt{\pi}}}\left( \begin{array}{c}
\alpha_1e^{i\gamma x} + \alpha_2e^{-i\gamma x}\\
-\sqrt{2}\alpha_1e^{i\gamma x} +\sqrt{2} \alpha_2e^{-i\gamma x}\\
\alpha_1e^{i\gamma x} + \alpha_2e^{-i\gamma x}\end{array}\right),
\label{initial_guess}
\end{equation}
where $|\alpha_1| = |\alpha_2| = 1/\sqrt{2}$ for $^{23}$Na and $|\alpha_1| = 1,|\alpha_2| = 0$
for $^{87}$Rb. Hence, by using different values of $|\alpha_1|$ and $|\alpha_2|$ in
 (\ref{initial_guess}), one can obtain different solutions corresponding to
the same density distribution and energy. For example, the two ground state solutions 
with ${\cal M} = 0$ for $^{23}$Na obtained by using $\alpha_1 = \alpha_2 = 1/\sqrt{2}$ 
and $\alpha_1 = 1/\sqrt{2}, \alpha_2 = -1/\sqrt{2}$ are shown in figures \ref{fig-1}(a) and
(b), respectively. In figures \ref{fig-1}(a) and (b), only the non-zero real (${\cal R}$) 
and imaginary (${\cal I}$) parts of the component wavefunctions are shown.
In these two cases, wavefunctions are either purely real or imaginary and not complex. 
On the other hand, the component wavefunctions in the ground state solution 
for $^{23}$Na obtained by using $\alpha_1 = 1/\sqrt{2}, \alpha_2 = i/\sqrt{2}$ 
are complex with non-zero real and imaginary parts. The real and imaginary parts of 
the component wavefunctions in this case are shown in figures \ref{fig-1} (c) and (d), respectively.
The multi-peak density profile corresponding to these three solutions presented in 
figures \ref{fig-1}(a),(b), and (c) and (d) is the  same and is shown in 
Fig. \ref{fig-1}(g). The multi-peak nature of the solution in this case is consistent 
with analytic results obtained in Sec. \ref{Sec-II}. The multi-peak solution effectively leads to a
weak phase separation between $\rho_{\pm 1}$ and $\rho_{0}$, here weak phase separation
implies that there are no local minima in the total density profile \cite{Ao}.
This is in contrast to the strong phase separation possible with the model
of the SO coupling discussed in Refs. \cite{gautam-1,gautam-2}, where a notch
appears in the total density profile at the interface separating the components
when $\gamma$ exceeds a critical value.
The solutions illustrated in  figures \ref{fig-1}(a),(b),  (c) and (d) are time-reversal symmetric. 
Similarly, the real and imaginary parts of the complex ground state solution 
with ${\cal M} = 0$ for $^{87}$Rb obtained with $\alpha_1 = 1/\sqrt{2}, \alpha_2 =  0$ 
in Eq. $(\ref{initial_guess})$ are shown in figures \ref{fig-1}(e) and (f), which lead to the 
single-peak density distribution  of figure \ref{fig-1}(h).  
The solution presented in figures \ref{fig-1}(e), (f), and (h) violates 
time-reversal symmetry, as there are two degenerate solutions in this case 
connected by the time-reversal operation.

\begin{figure}[!h]
\begin{center}
\includegraphics[trim = 0mm 6.5mm 0cm 0mm, clip,width=0.45\linewidth,clip]{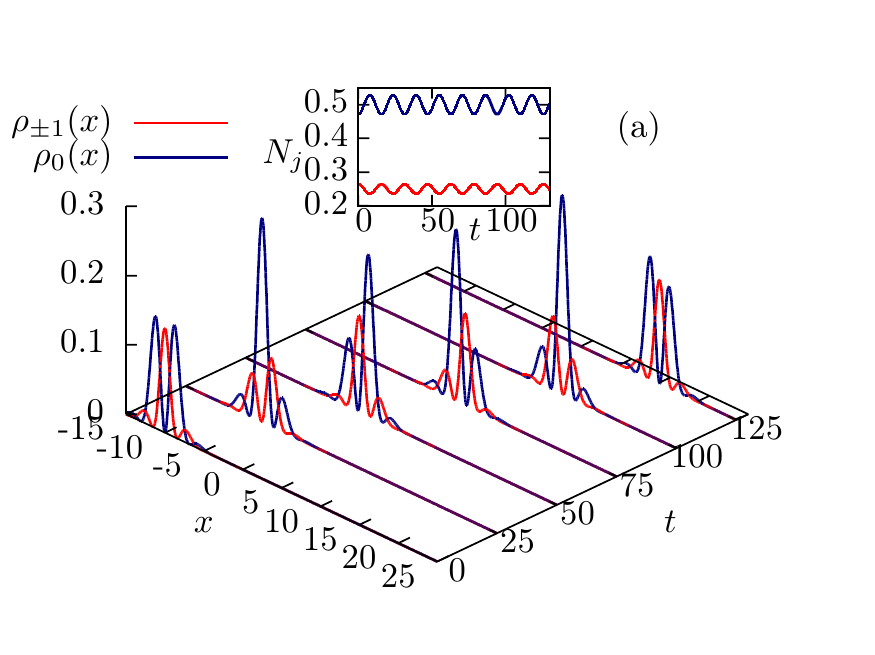}
\includegraphics[trim = 0mm 0mm 0cm 6.5mm, clip,width=0.45\linewidth,clip]{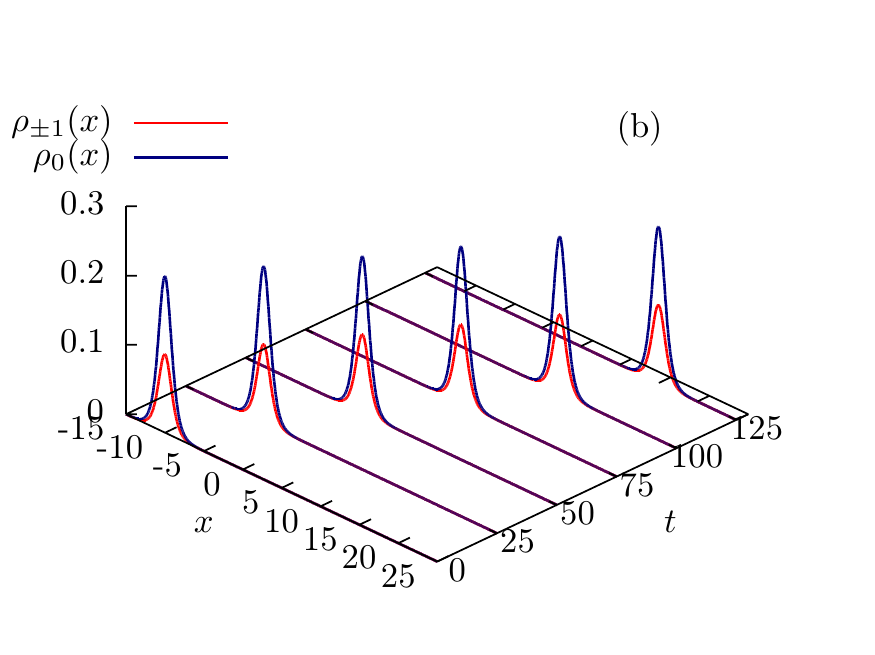}
\caption{(Color online) Propagation dynamics of the soliton of (a) figure \ref{fig-2} (a)
and (b)   figure \ref{fig-2} (b) in terms of component densities $\rho$. The inset
of (a) shows the variation in the population of the three components with time.  
At $t=0$, the solitons located at $x= -10$ are set into 
motion by multiplying the $t=0$ wave function by $\exp(i0.2x)$.}
\label{fig-3} \end{center}
\end{figure}

In order to obtain the bright solitons in SO-coupled spinor BECs, we take
$V(x)=0$ in  (\ref{gps-1})-(\ref{gps-3}) and consider  two
cases: (a) $c_0<0, c_2>0$ and (b) $c_0+c_2<0, c_2<0$. In case (a), we consider $c_0 = -1.2, c_2 = 0.3$. The numerically and variationally obtained bright solitons, defined
by  (\ref{ansatz_1}) and (\ref{sw1}) with ${\cal M}=0$, 
are shown in figures \ref{fig-2}(a). The multi-peak solution in this case  is time-reversal symmetric. In case (b), we consider  
$c_0 = -1.5, c_2 = -0.3$. The numerical and variational
solutions, defined by  (\ref{ansatz_2}) and (\ref{sw2}), in this case are shown 
in figure \ref{fig-2}(b). The single-peak  solution in this case breaks time-reversal 
symmetry of the Hamiltonian. It is evident from figure \ref{fig-2} that there is an 
excellent agreement between the numerical and variational results. 

In order to study the dynamics of the moving solitons, we first generate
the stationary solitons numerically using imaginary-time propagation for both
  antiferromagnetic and ferromagnetic interactions. In order to set these solitons into motion with
a constant velocity $v=0.2$, we multiply the wavefunction components for the stationary 
soliton with $\exp(i0.2x)$, and then use real-time propagation to study its evolution.
We observe that in the case of the antiferromagnetic soliton, there is spin-mixing dynamics due
to which the component densities are not conserved as the soliton moves. This is
evident from figure \ref{fig-3}(a) and its inset, which show the dynamics of the antiferromagnetic soliton
initially located at $x = -10$ and the spin-mixing dynamics, respectively; 
the interaction parameters are the same as those
in figure \ref{fig-2}(a) . At $t=0$ the soliton is set into motion at
a constant velocity.   As the soliton moves 
component densities keep on changing without any change in the total density. 
On the other hand, if one starts with the ferromagnetic soliton at $t=0$, the component
densities and hence the total density do not change while the soliton is moving.
This shown in figure \ref{fig-3}(b) for the soliton initially located at $x = -10$
and with the same interaction parameters as in figure \ref{fig-2}(b). 
This is consistent with the analytic results of Sec. \ref{Sec-III-B}.

\section{Summary}
\label{Sec-V}
We study the generation and propagation of a
 vector soliton with three components in an 
SO-coupled spin-1 BEC with either antiferromagnetic or ferromagnetic interactions.
In the antiferromagnetic case, the solutions  are time-reversal symmetric and 
the  component densities have multi-peak structure. In the ferromagnetic case, the solutions  violate  time-reversal symmetry and 
the  component densities have single-peak structure.
The GP equation for this system is not Galelian invariant.  From an analysis of the Galelian invariance of this equation, we establish that the single-peak ferromagnetic SO-coupled solitons can move with constant component densities and are true solitons, whereas the multi-peak  antiferromagnetic SO-coupled solitons change the component densities during motion.


\section*{Acknowledgements}
This work is financed by the Funda\c c\~ao de Amparo \`a Pesquisa do Estado de 
S\~ao Paulo (Brazil) under Contract Nos. 2013/07213-0, 2012/00451-0 and also by 
the Conselho Nacional de Desenvolvimento Cient\'ifico e Tecnol\'ogico (Brazil).

\end{document}